技術論文

# イオンペアクロマトグラフィー／電子スプレーイオン化質量分析法（LC/ESI-MS）によるアミノ酸のマススペクトル解析 *

高野 淑識 ** ・力石 嘉人 ** ・大河内 直彦 **

（2014 年 9 月 16 日受付，2014 年 12 月 1 日受理）

## 1. はじめに

　イオンペアクロマトグラフィーは，アミノ酸のように酸性および塩基性の双性イオンを有する有機分子の分離に有効である（例えば，Eksborg and Schill, 1973; Gloor and Johnson, 1977; Chaimbault et al., 2000）。移動相に適切なイオンペア剤，例えば，フッ素化アルキル基を有するカルボン酸等を添加することにより，目的化合物にイオンペアが形成されて，カラムへの保持が十分に行われる。このため，様々なイオンペア剤を任意に添加することは，界面活性効果による溶質の化学的コントロールも可能にして，計測したい有機分子の分離条件の最適化に寄与する（Knox and Hartwick, 1981; Cecchi, 2008）。このように，イオンペアクロマトグラフィーには，有機分子の分離に関して，順相および逆相クロマトグラフィーを補完できる利点がある。また，揮発性を有するイオンペア剤を選定することにより，ソフトイオン化を主目的とした電子スプレーイオン化質量分析計（ElectroSpray Ionization-Mass Spectrometry, ESI-MS; イオンソースの原理は，Yamashita and Fenn, 1984 および Appendix を参照）へ導入することができる。

　液体クロマトグラフィー／質量分析法（Liquid Chromatography/Mass Spectrometry, LC/MS）は，1970 年後半から 80 年代にその原理がプロトタイプ化されて以降（例えば，Arpino and Guiochon, 1979; Vestal, 1984; Covey et al., 1986），分離カラムの多様性が広がり，質量分析計の技術開発が併せて進んだことから，有機化学を主体とした幅広い分野で有用な分析手法になっている（例えば，大河内，2010; 佐藤，2011; 中村，2014 などの総説を参照）。さらに，目的成分に応じた前処理の検討（例えば，中村，2014）により，試料由来のマトリックス効果の軽減化，夾雑物溶出に伴うイオンサプレッション効果やエンハンス効果（例えば，Mallet et al., 2004; Antignac et al., 2005; Taylor, 2005）の最小化を図ることができれば，LC/MS の分析手法の利点を相乗的に発展・応用することができる。このような最適化を行い，例えば，極性官能基を有する機能性分子に着目した Kaneko et al. (2014) では，トリプル四重極質量分析計による多重反応モニタリング法（Multiple Reaction Monitoring, MRM）を応用し，0.1−1.0 femto mol（フェムトモル）オーダーのメタン生成および嫌気的メタン酸化に関わる補酵素分子の超高感度定量法を開発した。

　一方，LC/MS には，GC/MS のような統合マススペクトルライブラリー（例えば，NIST スペクトルデータベース：http://www.nist.gov/）が存在しない。このため，個々の LC/MS ユーザーが，各自の目的に合わせて，化合物標品から初期検討と最適化のすべてを行っている。このような，現在の

---

*LC/ESI-MS analysis of underivatized amino acids and mass spectrum
**国立研究開発法人 海洋研究開発機構 生物地球化学研究分野 〒 237-0061 横須賀市夏島町 2-15
　Yoshinori Takano, Yoshito Chikaraishi, Naohiko Ohkouchi
　Department of Biogeochemistry, Japan Agency for Marine-Earth Science and Technology (JAMSTEC), 2-15
　Natsushima, Yokosuka 237-0061, Japan
　Corresponding author: Yoshinori Takano
　E-mail: takano@jamstec.go.jp
　Fax: +81-46-867-9775

LC/MSを取りまく状況を少しでも前進させるために，重要な有機分子のマススペクトルデータを示しておくのは将来に有益であると考える。この技術論文では，イオンペアクロマトグラフィーを用いたアミノ酸の分離と質量分析によるマススペクトル解析を報告する。

## 2. 分析方法

### 2.1. アミノ酸試薬

アミノ酸は，試薬メーカー（Wako Pure Chemical Industries, Ltd.; Shanghai Hanhong Chemical Co., Ltd.; Sigma-Aldrich Co. LLC.）から入手したアミノ酸標品：グリシン（glycine），アラニン（L-α-alanine），セリン（L-serine），トレオニン（L-threonine），アスパラギン（L-asparagine），グルタミン（L-glutamine），アスパラギン酸（L-aspartic acid），グルタミン酸（L-glutamic acid），プロリン（L-proline），バリン（L-valine），リシン（L-lysine），ロイシン（L-leucine），イソロイシン（L-isoleucine），メチオニン（L-methionine），ヒスチジン（L-histidine），アルギニン（L-arginine），フェニルアラニン（L-phenylalanine），チロシン（L-tyrosine），ヒドロキシプロリン（L-hydroxyproline），サルコシン（sarcosine），β-アラニン（β-alanine），N-エチルグリシン（N-ethylglycine），α-アミノ酪酸（L-α-aminobutyric acid），α-アミノイソ酪酸（α-aminoisobutyric acid），β-アミノイソ酪酸（D,L-β-aminoisobutyric acid），γ-アミノ酪酸（γ-aminobutyric acid），イソバリン（L-isovaline），α-アミノアジピン酸（L-α-aminoadipic acid），ノルバリン（L-norvaline），ノルロイシン（L-norleucine）を用いた。各アミノ酸は，0.1M HClを用いてpH1に調整した純水に溶かして定容・調整した。タンパク性アミノ酸および非タンパク性アミノ酸の分子情報をTable 1に示す。

### 2.2. イオンペアクロマトグラフィーによるアミノ酸の分離と質量分析

アミノ酸は，誘導体化を行わず，液体クロマトグラフィー／電子スプレーイオン化質量分析法（HPLC/ESI-MS, Agilent 1100）を用いて分析した（Takano et al., 2015）。アミノ酸の導入量は，1.25 nmolから160 nmolの範囲で行った。分離カラムは，Hypercarb column（4.6×150 mm, particle size 5 μm; Thermo Fisher Scientific Inc.）および同ガードカラム（4.6×10 mm, particle size 5 μm）で行った。カラムおよびガードカラムは，カラムオーブン（Cool pocket column cooler, Thermo Fisher Scientific Inc.）で10.0℃に保持した。溶離液は，A：20 mMノナフルオロ吉草酸（Nonafluoropentanoic acid, NFPA；分子量 264.05），B：100％アセトニトリル（分子量 41.05）を用い，流速0.2 mL/minで，0分（B液：0％）から60分（B液：60％）のグラジエントで行い，60–70分でバックフラッシュを行った後，30分（B液：0％）の平衡化時間を確保した。分離カラムの固定相とイオンペア剤の親和性を確保するために，分析の前日に一晩，同条件かつ流速0.1 mL/minでカラムの初期平衡化を行った。分析前と分析後の十分な平衡化の時間は，分離および保持時間の安定性を確保するために重要である。イオンペア剤の最適化は，既報（例えば，Cecchi, 2008）に詳しくまとめられている。

アミノ酸の検出は，前述のオンライン分離で電子スプレーイオン化質量分析計（ESI, positive mode）へ導入し，質量範囲 $m/z$ 70–400で行った。イオンソースの条件は，ドライガス窒素温度200℃，ドライガス窒素流量10 L，ネブライザー圧力50 psi，キャピラリー電圧3000Vで行った。ポジティブイオンモードおよびネガティブイオンモードでの適用について，溶質と溶液に関する酸塩基理論の原理をFigure 1に示す。ESIで生成するイオンは，試料溶液のpHに依存する傾向があるので，試料のコンディショニングは重要である。ここで，アミノ酸はイオン化され，$[M+H]^+$の形で検出される。光学検出器は，ダイオードアレイ検出器（DAD, Agilent 1100）を用い，吸光特性のある芳香族アミノ酸を検出し，アミノ酸の分離状態とカラムのコンディショニングをモニターした（220–400 nmを検出；Appendix）。

## 3. 結果と考察

### 3.1. アミノ酸のマスフラグメントパターン

イオンペアクロマトグラフィーによる各アミノ酸の分離例をFigure 2に，アミノ酸の代表的なマスフラグメントパターンをFigure 3に示す。前述のように，pHを酸性側に調整したアミノ酸を用い

**Table 1.** Summary of the ion-pair reversed-phase LC for underivatized amino acids showing elution order, chemical formula, molecular weight, parent ion, and fragment (*m/z*) by electrospray ionization mass spectrometry. #1, hydroxyproline is one of the important amino acids in collagen protein. #2, #3, Asparagine (Asn) and glutamine (Gln) will convert to aspartic acid (Asp) and glutamic acid (Glu), respectively, after hydrolysis. #4, The chromatographic co-elution of leucine and isoleucine may occur on this ion-pair LC separation. However, if the eluent and gradient program was modified with same ion-pair reagent, leucine and isoleucine were separated as shown in Takano et al. (2015) and the supplementary information. In the right-hand column, 'P' and 'NP' represent protein amino acid and non-protein amino acid, respectively.

| Elution order | Abbreviation | Formula | Molecular weight | Retention time (min) | Parent ions $[M+H]^+$ | Product ions (*m/z*) | Remarks |
|---|---|---|---|---|---|---|---|
| Glycine | Gly | $C_2H_5NO_2$ | 75 | 18.6 | 76 | – | P |
| Serine | Ser | $C_3H_7NO_3$ | 105 | 22.3 | 106 | 88 | P |
| Sarcosine | Sar | $C_3H_7NO_2$ | 89 | 23.6 | 90 | – | NP |
| Alanine | Ala | $C_3H_7NO_2$ | 89 | 25.2 | 90 | – | P |
| Hydroxyproline | Hyp | $C_5H_9NO_3$ | 131 | 26.3 | 132 | – | NP (#1) |
| $\beta$-Alanine | $\beta$-Ala | $C_3H_7NO_2$ | 89 | 26.4 | 90 | 72 | NP |
| Threonine | Thr | $C_4H_9NO_3$ | 119 | 27.0 | 120 | 102, 74 | P |
| Asparagine | Asn | $C_4H_8N_2O_3$ | 132 | 27.3 | 133 | 116, 87 | P (#2) |
| N-Ethylglycine | N-Et-gly | $C_4H_9NO_2$ | 103 | 28.3 | 104 | – | NP |
| Glutamine | Gln | $C_5H_{10}N_2O_3$ | 146 | 29.3 | 147 | 130, 101 | P (#3) |
| $\beta$-Aminoisobutyric acid | $\beta$-AiBA | $C_4H_9NO_2$ | 103 | 29.5 | 104 | – | NP |
| Aspartic acid | Asp | $C_4H_7NO_4$ | 133 | 29.5 | 134 | 116, 88 | P |
| $\alpha$-Aminoisobutyric acid | $\alpha$-AiBA | $C_4H_9NO_2$ | 103 | 29.6 | 104 | – | NP |
| $\gamma$-Aminobutyric acid | $\gamma$-ABA | $C_4H_9NO_2$ | 103 | 30.1 | 104 | – | NP |
| Proline | Pro | $C_5H_9NO_2$ | 115 | 30.1 | 116 | – | P |
| $\alpha$-Aminobutyric acid | $\alpha$-ABA | $C_4H_9NO_2$ | 103 | 30.4 | 104 | – | NP |
| Glutamic acid | Glu | $C_5H_9NO_4$ | 147 | 31.9 | 148 | 130, 102 | P |
| Isovaline | Isoval | $C_5H_{11}NO_2$ | 117 | 34.4 | 118 | 72 | NP |
| Valine | Val | $C_5H_{11}NO_2$ | 117 | 34.8 | 118 | 72 | P |
| $\alpha$-Aminoadipic acid | $\alpha$-AAA | $C_6H_{11}NO_4$ | 161 | 35.4 | 162 | 144, 116 | NP |
| Norvaline | Norval | $C_5H_{11}NO_2$ | 117 | 35.7 | 118 | 72 | NP |
| Lysine | Lys | $C_6H_{14}N_2O_2$ | 146 | 36.6 | 147 | 130 | P |
| Leucine | Leu | $C_6H_{13}NO_2$ | 131 | 38.5 | 132 | 86 | P |
| Isoleucine | Ile | $C_6H_{13}NO_2$ | 131 | 38.5 | 132 | 86 | P |
| Methionine | Met | $C_5H_{11}NO_2S$ | 149 | 39.0 | 150 | 133, 104 | P |
| Histidine | His | $C_6H_9N_3O_2$ | 155 | 40.1 | 156 | 110 | P |
| Norleucine | Norleu | $C_6H_{13}NO_2$ | 131 | 40.4 | 132 | 86 | NP |
| Arginine | Arg | $C_6H_{14}N_4O_2$ | 174 | 43.1 | 175 | – | P |
| Phenylalanine | Phe | $C_9H_{11}NO_2$ | 165 | 51.3 | 166 | 120 | P |
| Tyrosine | Tyr | $C_9H_{11}NO_3$ | 181 | 53.2 | 182 | 165, 136 | P |

**(a) Positive ion mode (pH < 7)**

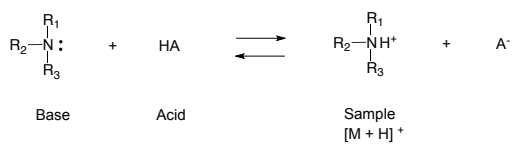

**(b) Negative ion mode (pH > 7)**

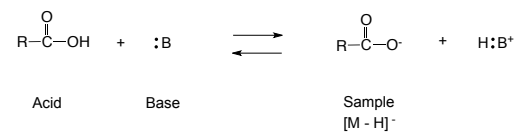

**Fig. 1.** The principles of acid-base theory in solution chemistry for LC/ESI-MS optimization.

た ESI-MS のポジティブモードでは，$[M+H]^+$ の親イオン（プロトン付加分子イオン）に加え，アミノ酸の官能基部位で開裂した後，H 原子が付加したプロダクトイオンが生成する。

まず，アミノ酸のマススペクトルには，カルボキシル基（-COOH）の脱離による生成したイオン$[M+H-HCOOH]^+=[M+H-46]^+$ が見られる分子種がある（Figure 3-a）。例えば，バリンのマススペクトルは，$[M+H]^+=118$ であり，$[M+H-46]^+$ に対応する *m/z* 72 がある（Figure 4）。このフラグメントパターンは，アルキル基を有するアミノ酸であるノルバリン，イソバリン，ロイシン，ノルロイシン，芳香族アミノ酸であるフェニルアラニン，チロシン，複素環式アミノ酸であるヒスチジンに

Fig. 2. (1/2)

**(a) Protein AAs**

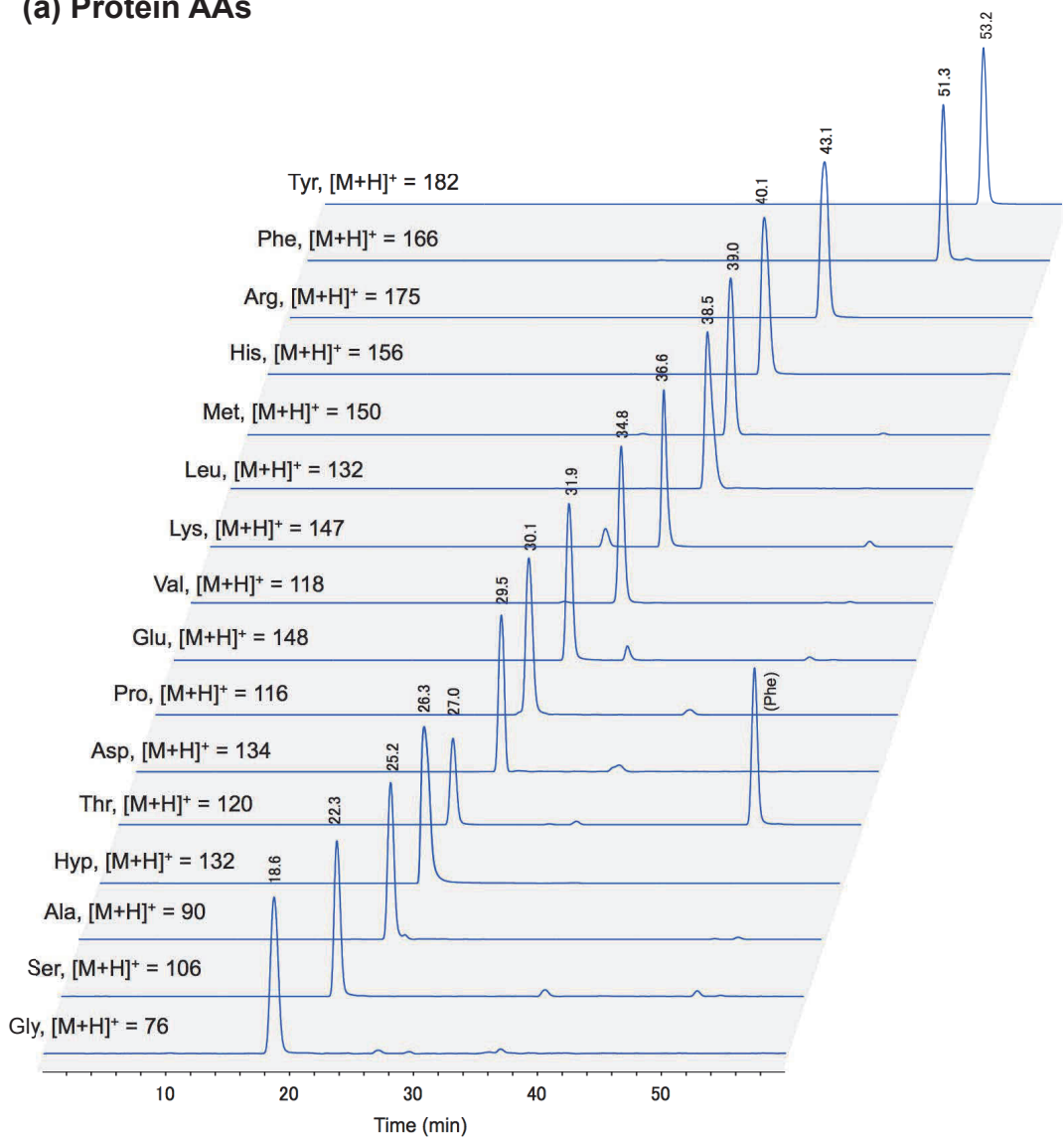

**Fig. 2.** Extracted ion chromatogram (EIC) of underivatized amino acids on LC/ESI–MS analysis. Abbreviations: Glycine, Gly; Serine, Ser; Alanine, Ala; Threonine, Thr; Aspartic acid, Asp; Proline, Pro; Glutamic acid, Glu; Valine, Val; Lysine, Lys; Leucine, Leu; Isoleucine, Ile; Methionine, Met; Histidine, His; Arginine, Arg; Phenylalanine, Phe; Tyrosine, Tyr; Asparagine, Asn; Glutamine, Gln; Sarcosine, Sar; Hydroxyproline, Hyp; *β*-Alanine, *β*-Ala; N-Ethylglycine, N-Et-gly; *β*-Aminoisobutyric acid, *β*-AiBA; *α*-Aminoisobutyric acid, *α*−AiBA; *γ*-Aminobutyric acid, *γ*-ABA; *α*-Aminobutyric acid, *α*-ABA; Isovaline, Isoval; *α*-Aminoadipic acid, *α*-AAA; Norvaline, Norval; Norleucine, Norleu. A co-injection of threonine and phenylalanine showed the same product ion (*m/z* 120) at differing retention times. For separation condition, please see Takano et al. (2015).

Fig. 2. (2/2)

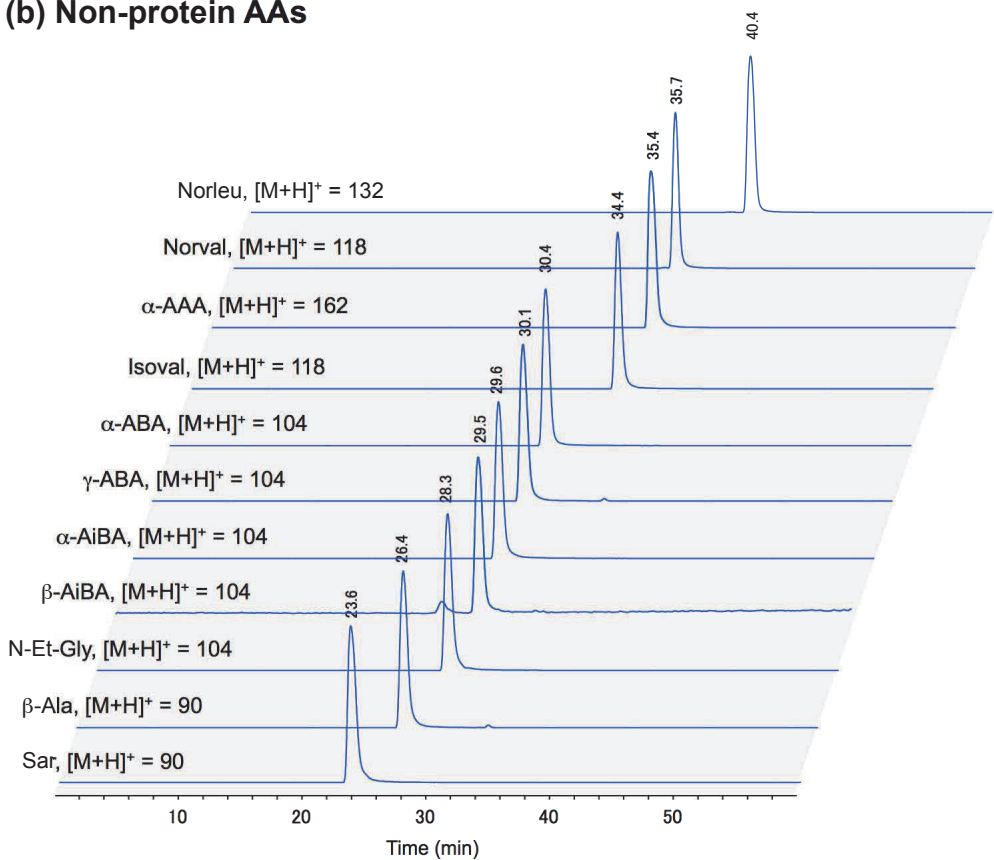

**(b) Non-protein AAs**

**(c) Unhydrolyzed protein AAs**

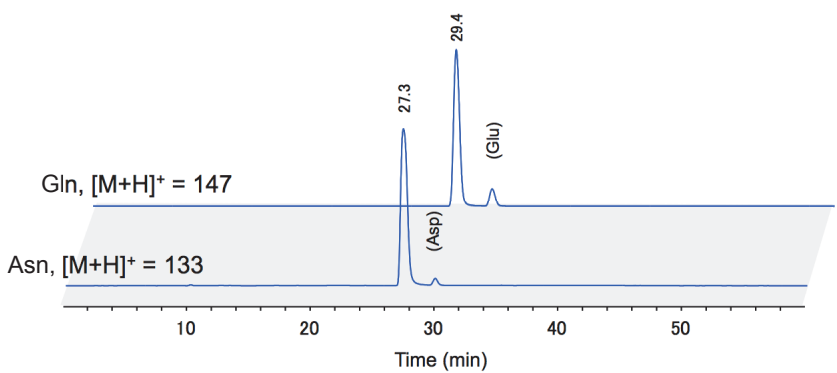

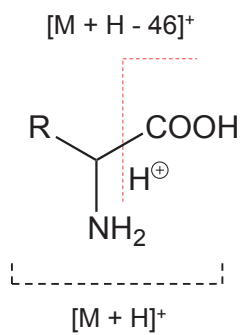

**(a) -COOH fragment**

$[M+H-46]^+$

$[M+H]^+$

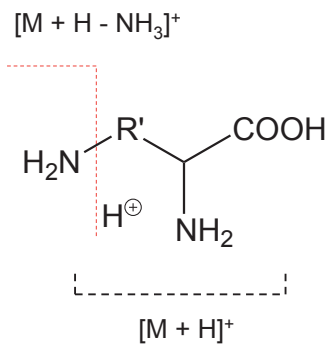

**(b) -NH$_2$ fragment**

$[M+H-NH_3]^+$

$[M+H]^+$

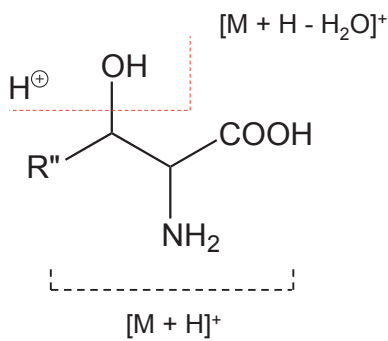

**(c) -OH fragment**

$[M+H-H_2O]^+$

$[M+H]^+$

**Fig. 3.** Representative fragmentation patters for diagnosis ion of underivatized amino acids showing, parent ion $[M+H]^+$ with **(a)** –COOH fragment (-HCOOH), **(b)** –NH$_2$ fragment (-NH$_3$), and **(c)** –OH fragment (-H$_2$O).

共通している．アミノ基を2つ有するグルタミンやアスパラギン等にも見られる．

次に，アミノ基（-NH$_2$）の脱離による生成したイオン $[M+H-NH_3]^+$ = $[M+H-17]^+$ が見られる分子種もある（Figure 3-b）．例えば，リシンのマススペクトルは，$[M+H]^+$ = 147 であり，$[M+H-17]^+$ に対応する $m/z$ 130 がある（Figure 4）．このフラグメントパターンは，アスパラギンやグルタミン，あるいは，含硫アミノ酸であるメチオニン等にも共通する．

ヒドロキシル基（-OH）の脱離による生成したイオン $[M+H-H_2O]^+$ = $[M+H-18]^+$ が見られる場合もある（Figure 3-c）．例えば，セリンのマススペクトルは，$[M+H]^+$ = 106 であり，$[M+H-18]^+$ に対応する $m/z$ 88 がある（Figure 4）．このフラグメントパターンは，ヒドロキシル基を有するアミノ酸であるトレオニン，酸性アミノ酸のアスパラギン酸やグルタミン酸，あるいは，α-アミノアジピン酸や β-アラニンにも共通する．

**3.2. 本質量分析の留意点と他の検出器への応用**

各アミノ酸の保持時間とマススペクトルを比較することにより，アミノ酸の同定を行うことができ，さらに，濃度既知のスタンダードを用いれば，定量を行うことができる．ここで挙げているプロダクトイオンは，前述の質量分析計の条件で行ったものであり，例えば，イオン化の設定条件を変えることにより，フラグメントパターンにも多少の変化が現れることがある．前述 3.1. のような主要イオンの他，例えば，グルタミン酸では，$[M+H-HCOOH-H_2O]^+$ が生じ，グルタミンでは，$[M+H-HCOOH-NH_3]^+$ が生じる．

トレオニン（$[M+H]^+$ = 120）とフェニルアラニン（$[M+H]^+$ = 166）を共打ちすると，フェニルアラニンのカルボキシル基由来のフラグメント化により（$[M+H-46]^+$ = 120），同じイオンクロマトグラム（EIC, $m/z$ 120）上にシグナルが検出される（Figure 2）．質量数が同一で保持時間が近いアミノ酸分子が含まれる試料の場合，後述の2次元分析によるクロスチェックが重要である．タンパク性および非タンパク性アミノ酸の溶出順序，略記，化学式，分子量，プロダクトイオンを Table 1 にまとめた．ここで用いているノナフルオロ吉草酸の

Fig. 4. (a) (1/5)

## (a) Protein AAs

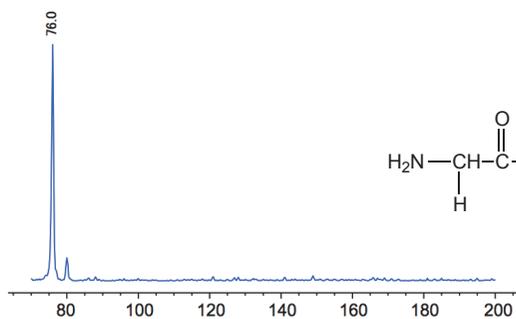
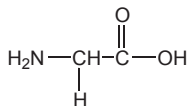

Gly, [M+H]$^+$ = 76

Chemical Formula: $C_2H_5NO_2$
Exact Mass: 75.03
Molecular Weight: 75.07

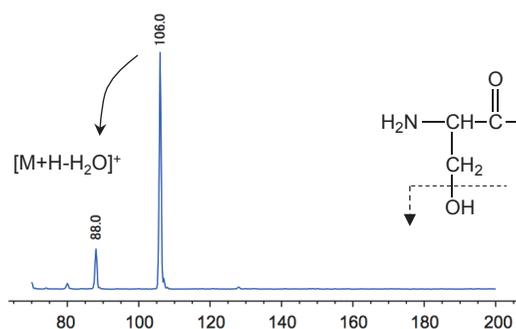
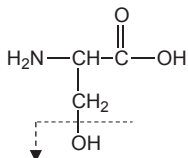

Ser, [M+H]$^+$ = 106

Chemical Formula: $C_3H_7NO_3$
Exact Mass: 105.04
Molecular Weight: 105.09

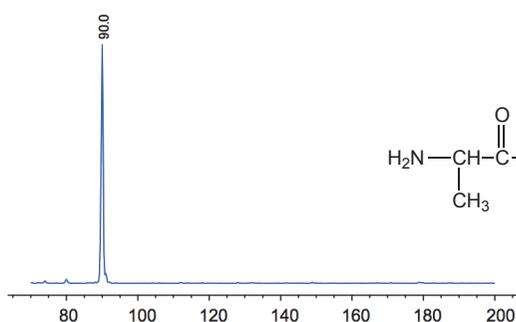
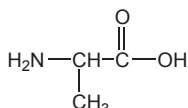

Ala, [M+H]$^+$ = 90

Chemical Formula: $C_3H_7NO_2$
Exact Mass: 89.05
Molecular Weight: 89.09

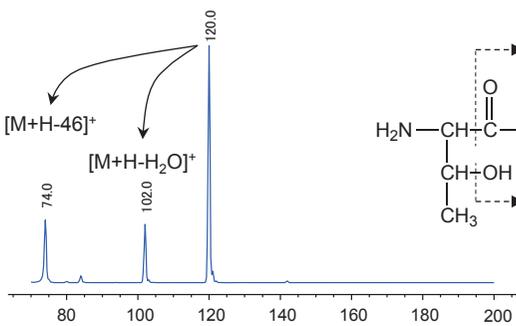
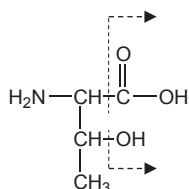

Thr, [M+H]$^+$ = 120

Chemical Formula: $C_4H_9NO_3$
Exact Mass: 119.06
Molecular Weight: 119.12

*m/z*

**Fig. 4. (a)** Underivatized protein amino acids and **(b)** non-protein amino acids for the identification. Observed product ions and their fragmentation patterns by ESI–MS were shown. The chemical formula, exact mass, molecular weight, and theoretical values of their corresponding masses (*m/z*) are also shown.

Fig. 4. (a) (2/5)

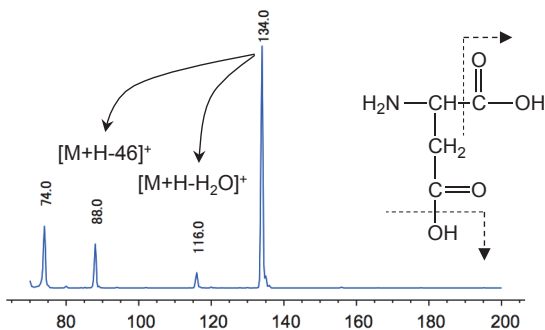

Asp, [M+H]$^+$ = 134

Chemical Formula: $C_4H_7NO_4$
Exact Mass: 133.04
Molecular Weight: 133.10

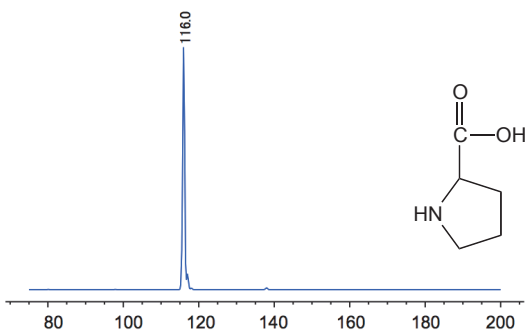

Pro, [M+H]$^+$ = 116

Chemical Formula: $C_5H_9NO_2$
Exact Mass: 115.06
Molecular Weight: 115.13

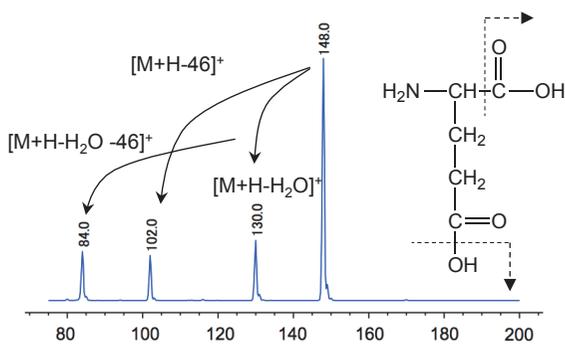

Glu, [M+H]$^+$ = 148

Chemical Formula: $C_5H_9NO_4$
Exact Mass: 147.05
Molecular Weight: 147.13

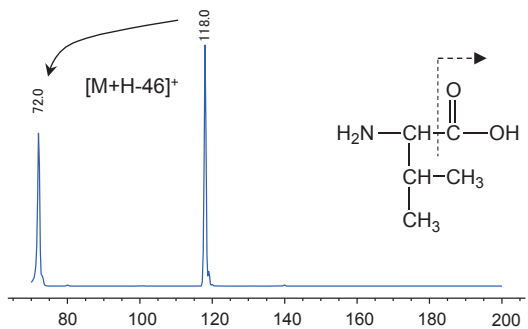

Val, [M+H]$^+$ = 118

Chemical Formula: $C_5H_{11}NO_2$
Exact Mass: 117.08
Molecular Weight: 117.15

Fig. 4. (a) (3/5)

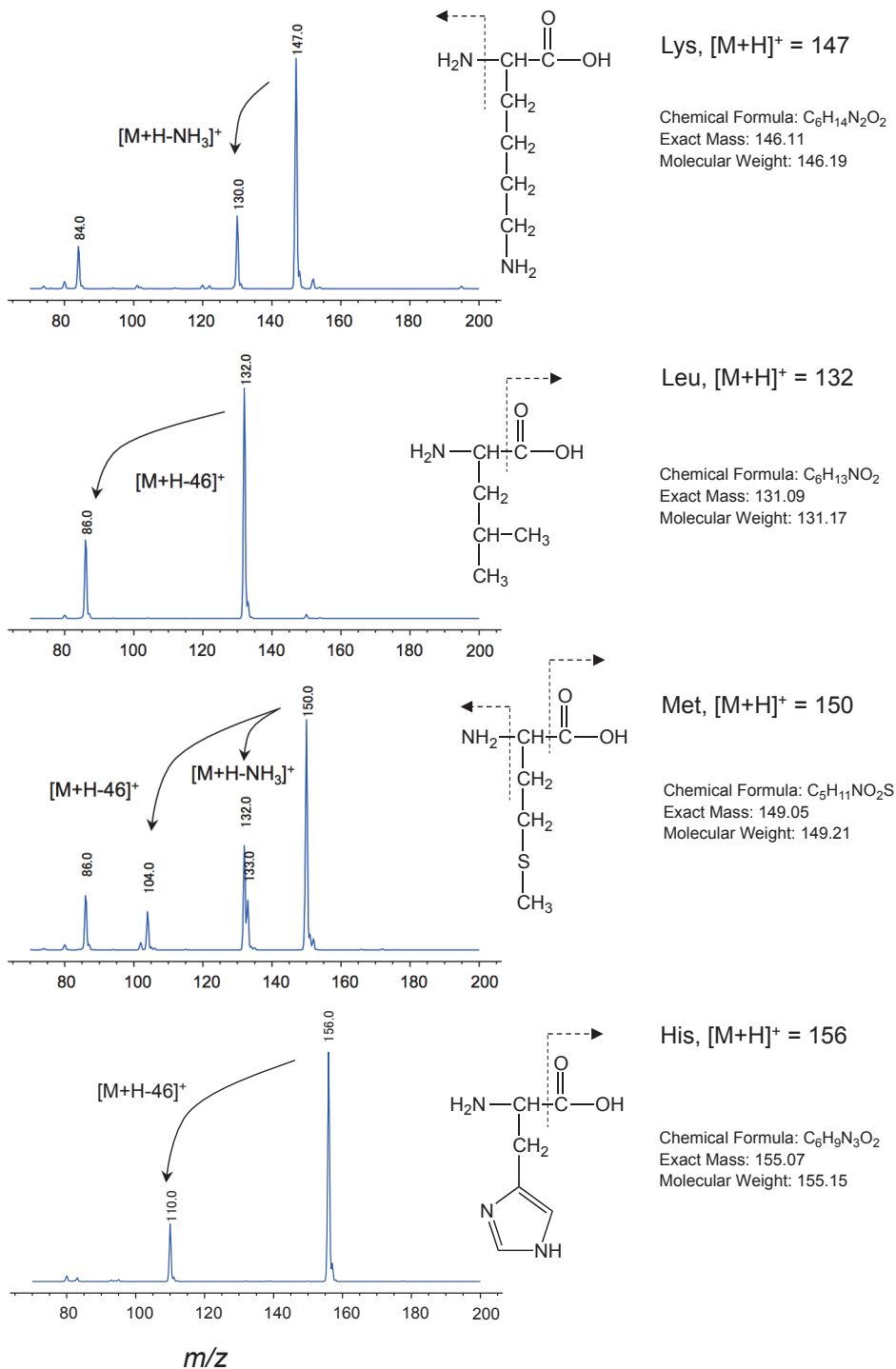

Fig.4. (a) (4/5)

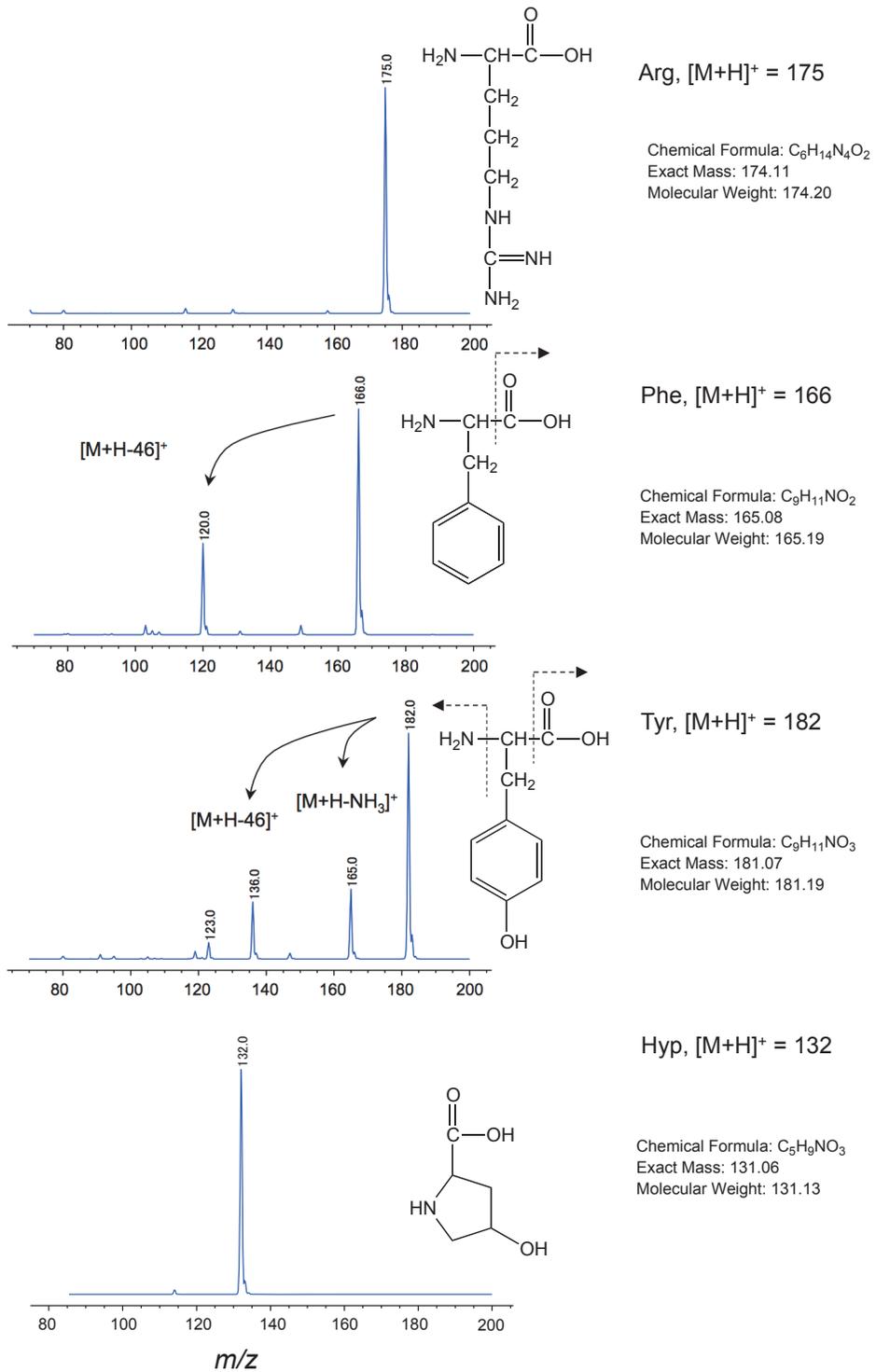

Fig. 4. (a) (5/5)

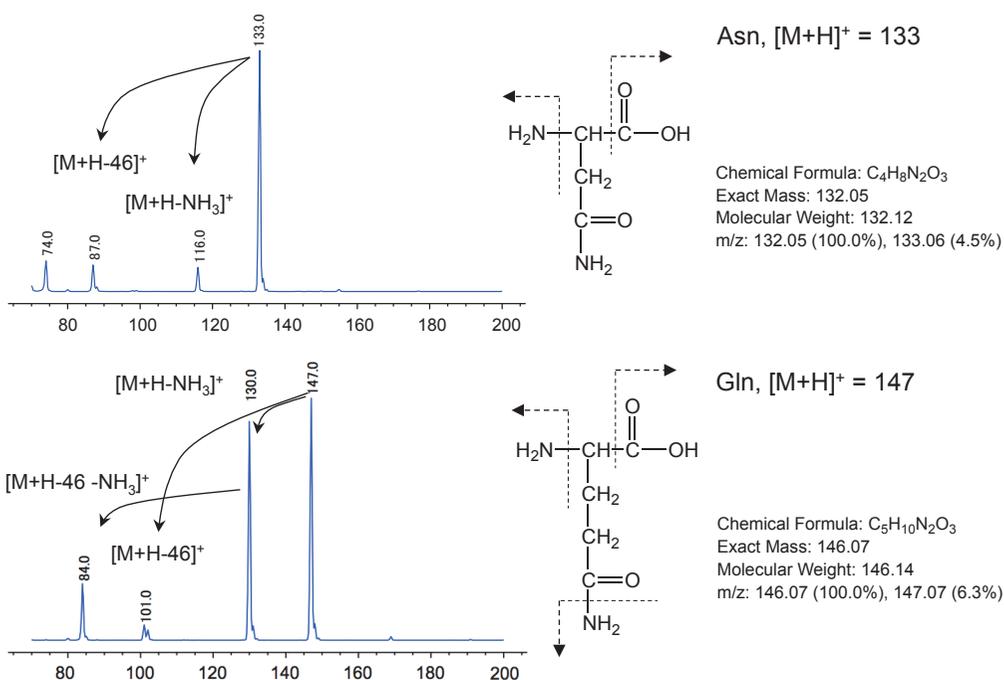

分子量は，264.05 であり，Table 1 に示すアミノ酸分子のマススペクトルに干渉しない。

アミノ酸の同定について十分な確認が行われている場合，質量分析計の代わりにコロナ荷電化粒子検出器（Corona CAD, Charged Aerosol Detector）や蒸発光散乱検出器（Evaporative Light Scattering Detector, ELSD）による検出を行うことが可能である。Corona CAD（Appendix: Thermo Fisher Scientific Inc.）は，定量性に優れ，日常のメンテナンスが質量分析計よりも比較的容易であることから，質量分析法やフォトダイオードアレイ検出法の補完的な検出装置として有効と考えられる。

### 4. 今後の展望

**4.1. 二次元クロマトグラフィーへの応用**

有機分子を1次元LCで分離し，必要性に応じて2次元目のクロマトグラフィーへの展開することが可能である。2次元クロマトグラフィーへのアプローチは，非タンパク性アミノ酸とタンパク性アミノ酸が混在している特殊な試料を測定する際の分析確度保証にも有効である。Hamase et al.（2014）では，非タンパク性アミノ酸を含む網羅的な立体異性体比（D/L比）の正確な記載を報告している。また，アミノ酸に揮発性を持たせるために，適切な誘導体化（例えば，N-ピバロイルイソプロピル化；力石ら，2009；エトキシカルボニル-エチルエステル化；山口ら，2009）およびガスクロマトグラフィーによる分離（例えば，Chikaraishi et al., 2010）を経て，包括的 LC x GC，あるいは，マルチハートカット LC-GC へと応用することも可能となる。すなわち，試料由来の網羅的なアミノ酸組成（例えば，Chikaraishi et al., 2009）から，任意の分子レベルで，より高分解能なターゲット分析へ進めることを意味する。

**4.2. アミノ酸分子レベル同位体研究への展望**

本稿で述べたアミノ酸の分離は，誘導体化を行っていない。著者らは，アミノ酸のアミノ基由来の窒素について，前処理と分離の際に同位体分別が無い分析条件の最適化を行ってきた（例えば，Chikaraishi et al., 2010; Takano et al., 2010, 2015）。こ

Fig. 4. (b) (1/3)

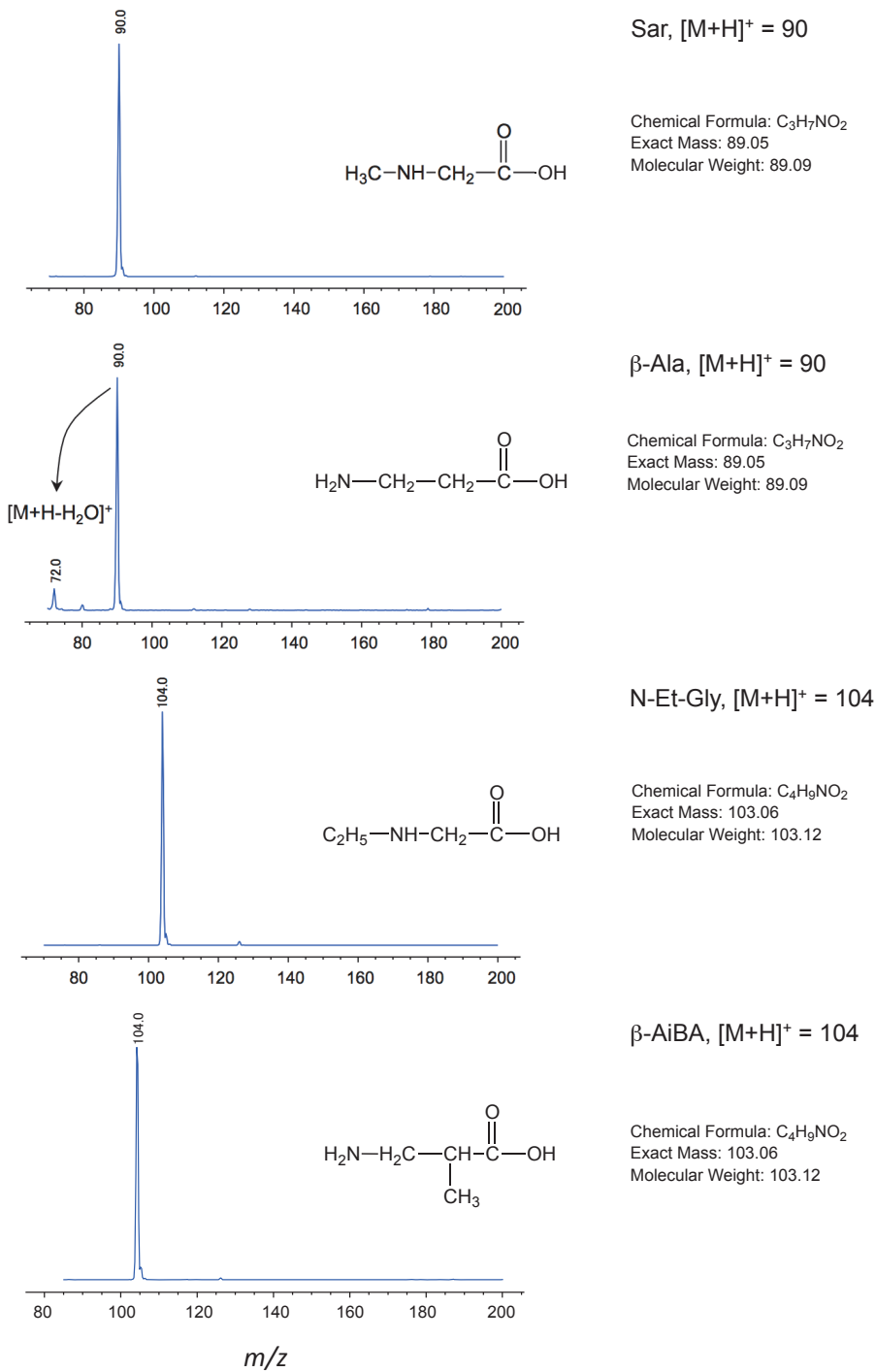

**Fig. 4. (a)** Underivatized protein amino acids and **(b)** non-protein amino acids for the identification. Observed product ions and their fragmentation patterns by ESI–MS were shown. The chemical formula, exact mass, molecular weight, and theoretical values of their corresponding masses ($m/z$) are also shown.

Fig. 4. (b) (2/3)

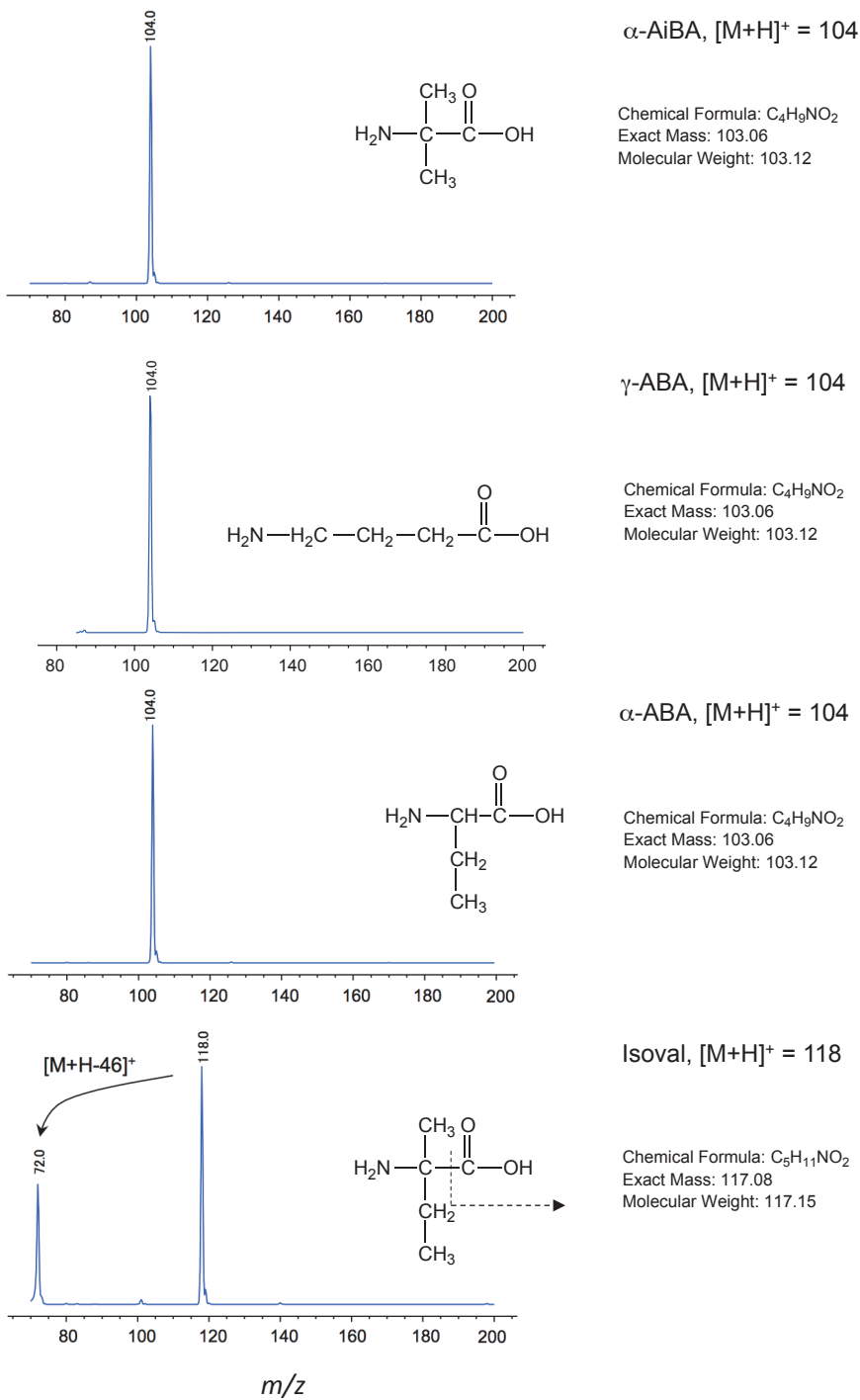

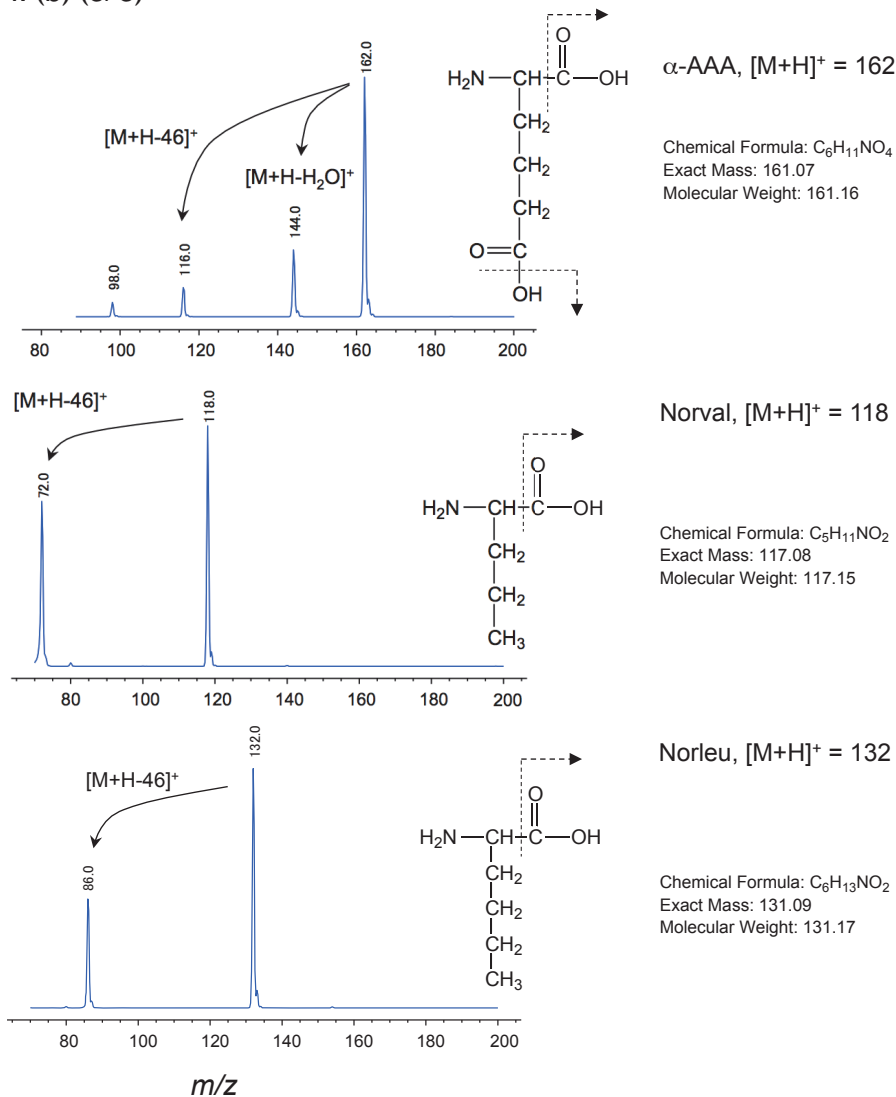

Fig. 4. (b) (3/3)

のことは，マトリックス効果を効率的に除去することでアミノ酸分子レベルあるいは立体異性体分子レベル（Takano et al., 2009; Ohkouchi and Takano, 2014）での窒素同位体比の挙動を高精度かつ高確度で記載できることを意味する。ここで示した既報では，アミノ酸の炭素同位体比の検証（例えば，Chikaraishi and Ohkouchi, 2010）については触れていないが，分離条件と精製法の最適化を行うことにより，極微量スケール（例えば，Ogawa et al., 2010）での炭素同位体比（$\delta^{13}C$）の動態を含め，アミノ酸分子レベル放射性炭素年代測定（例えば，Itahashi et al., 2014）への展開に道が開かれる。

### 謝　辞

**Abbreviations :** LC/MS, Liquid Chromatography/Mass Spectrometry; ESI-MS, ElectroSpray Ionization-Mass Spectrometry; EIC, Extracted Ion Chromatogram; DAD, Diode Array Detector; MRM, Multiple Reaction Monitoring; NIST, National Institute of Standards and Technology; Corona CAD, Corona Charged Aerosol Detector; ELSD, Evaporative Light Scattering Detector.

イオンペアクロマトグラフィー／電子スプレーイオン化質量分析法
（LC/ESI-MS）によるアミノ酸のマススペクトル解析

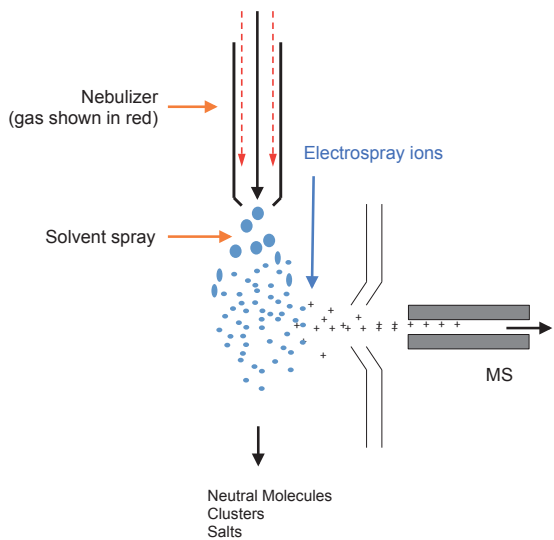

**appendix. 1.** Illustration of typical electrospray source (ESI).

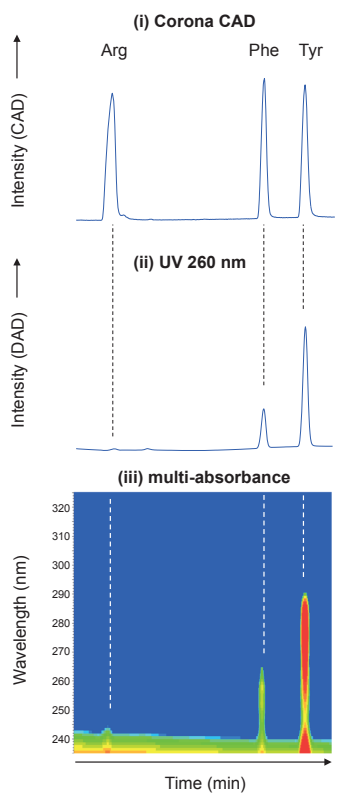

**appendix. 2.** Comparison of responses determined by other online detectors for corona CAD (Thermo Fisher Scientific Inc.) and the diode array detector (DAD) with the UV absorbance. Modified after Takano et al. (2015)

― 17 ―